\documentclass[journal]{IEEEtran}

\usepackage{cite}
\usepackage{amsmath,amssymb,amsfonts}
\usepackage{algorithmic}
\usepackage{graphicx,color}
\usepackage{textcomp}
\usepackage{hyperref}

\hypersetup{
    colorlinks = true,
    linkcolor  = black,   
    citecolor  = blue,    
    urlcolor   = blue     
}

\newif\ifreview

\reviewfalse   

\newcommand{\rev}[1]{%
\ifreview
\textcolor{blue}{#1}
\else
#1
\fi
}

\usepackage{cleveref}
\crefname{figure}{Fig.}{Figs.}
\Crefname{figure}{Figure}{Figures}

\usepackage{xcolor}

\usepackage[caption=false,font=footnotesize]{subfig}

\title{Ray-Based Simulation of Scattering from Discretized Curved Bodies for Vehicular and  ISAC Applications}

\author{
\IEEEauthorblockN{
Ainur Ziganshin\IEEEauthorrefmark{1},
Enrico M. Vitucci\IEEEauthorrefmark{2},
Wim Kotterman\IEEEauthorrefmark{1},
Reiner Thom\"a\IEEEauthorrefmark{1},\\
Christian Schneider\IEEEauthorrefmark{1},
Vittorio Degli-Esposti\IEEEauthorrefmark{2}
}

\IEEEauthorblockA{\IEEEauthorrefmark{1}
Electronic Measurements and Signal Processing Group (EMS) at Th\"uringer Innovationszentrum Mobilit\"at, Technische Universit\"at Ilmenau, Ilmenau, Germany}

\IEEEauthorblockA{\IEEEauthorrefmark{2}
Dipartimento dell’Ingegneria Elettrica e dell’Informazione (DEI),
Universit\`a di Bologna, Bologna, Italy}

\thanks{Corresponding author: Ainur Ziganshin (ainur.ziganshin@tu-ilmenau.de).}
\thanks{This is a preprint version of a manuscript submitted to the IEEE Open Journal of Antennas and Propagation.}
}

\begin{document}

\maketitle

\begin{abstract}
Realistic modeling of scattering from curved metallic bodies — such as vehicles and roadside structures — is essential for cellular and vehicular channel modeling as well as radar applications. A practical approach is to approximate curved surfaces with planar facets and apply ray-tracing with diffraction methods; however, accuracy depends critically on both geometric discretization and diffraction modeling.
\rev{This work investigates ray-tracing-based modeling of near-field scattering from curved bodies, both in the backscattering and in the forward (shadow) region; in the ray-tracing tool, diffraction is modeled according to the Uniform Theory of Diffraction (UTD), extended with vertex diffraction and double-bounce interactions, including a heuristic combination of edge and vertex diffraction.} A discretization strategy linking facet size to local curvature and wavelength is proposed to balance geometric fidelity, diffraction modeling, and efficiency. 
Validation is initially performed against analytical solutions and full-wave simulations for canonical geometries (sphere and circular cylinder). 
\rev{Furthermore, the practical applicability of the approach is demonstrated for a realistic vehicle by comparison with bistatic measurements in the backscattering region and full-wave simulation in the shadow region.}
The results demonstrate that no universal discretization strategy exists: \rev{fine meshes are beneficial for accurate backscattering prediction, while coarser discretizations can provide more efficient and accurate shadow region prediction.}
The proposed extended diffraction framework provides a computationally efficient framework for vehicular propagation and integrated sensing and communication (ISAC) channel modeling.
\end{abstract}


\section{Introduction}

Vehicular applications are becoming a cornerstone of next-generation wireless systems, including vehicle-to-everything (V2X) communications, integrated sensing and communication (ISAC), and automotive radar, \cite{thoma2025distributed, noor20226g}. 
In these scenarios, radio propagation is strongly influenced by the presence of vehicles themselves, which act as large, mobile, and often dominant scatterers in the environment. 
Moreover, in ISAC and radar applications, the target-induced channel components, characterized by their delay, Doppler, and angular signatures, constitute the main focus. 

Accurate and efficient modeling of vehicle-related scattering and blockage is therefore essential for large-scale simulation, digital twins, system optimization, and testing under realistic conditions.
These applications typically deal with electrically large objects, wide frequency ranges, and multiple scenarios, and therefore require low computational effort with satisfactory accuracy at the same time \cite{ziganshin2024scalable}.

A key challenge in propagation modeling in the presence of vehicles is the accurate representation of scattering and blockage caused by curved bodies. 
Unlike buildings, which are often approximated as planar wall structures, vehicles exhibit smooth and moderately curved metallic surfaces, making modeling more challenging.
Moreover, vehicular scenarios frequently involve multi-static configurations and near-field propagation conditions, further complicating the modeling procedure, so that simplified approaches based on the standard, far-field Radar Cross Section (RCS) concept are often inadequate.

\rev{Vehicle scattering has been extensively investigated through both experimental and numerical approaches. Measurement campaigns \cite{andrich2026bira}, including RCS and bistatic reflectivity studies, have provided valuable insights into the scattering behavior of different vehicle classes and supported the development of statistical models \cite{myint2019statistical}.}
In parallel, high-frequency electromagnetic techniques \cite{davidson2010computational}, \cite{gibson2021method} such as the finite-difference time-domain (FDTD) method,  integral-equation solvers like the method of moments (MoM), or physical optics (PO) with physical theory of diffraction (PTD) \cite{ufimtsev2014fundamentals} have been widely used to simulate scattering from realistic geometries with high accuracy \cite{weinmann2006ray,18706}. 
These approaches are well-suited for detailed electromagnetic analysis and often serve as reference solutions; however, their computational cost typically limits their applicability in large-scale wireless propagation studies.

\rev{Ray Tracing (RT), in the form of forward-ray-tracing or \emph{Shooting and Bouncing Rays} (SBR), has been applied to the accurate calculation of the RCS of electrically large and complex aircraft bodies since the latter decades of the past century. In this hybrid approach, SBR tracks ray trajectories and multiple reflections on the aircraft surface, while PO integration of the induced surface currents yields an accurate computation of the backscattered field and thus the RCS \cite{18706, weinmann2006ray}, and references therein. This SBR+PO approach, however, is limited to the illuminated region and is not suitable to predict the scattered field in the shadow region of the obstacle. Furthermore, the need to cascade PO after RT negates the computational advantages of RT.}

Motivated by the need for scalable simulation of a broader range of cases and distances, ray-tracing has subsequently become a standard tool for wireless channel modeling in urban, indoor, and vehicular environments, where high accuracy is not required \cite{7152831}.
However, very few investigations have applied pure ray-based techniques to the multistatic solution of near-field reflectivity problems encountered in vehicular applications \cite{4685913, sadovaya2020ray, gou2024improved}. This is due in part to the difficulty in treating diffraction in discrete surface representations using small facets and edges, as explained below.

While curved surfaces can be treated directly using parametric representations such as Non-Uniform Rational B-Splines (NURBS) \cite{67702}, and several studies have investigated ray-tracing over smooth parametric surfaces \cite{domingo1995computation, 6512614, sefi2003pay}, these approaches are computationally demanding and are not commonly supported by practical RT frameworks.
Consequently, reference RT tools, such as the open-source, parallelizable framework Sionna-RT \cite{hoydis2023sionna} and other tools \cite{eertmans2025fast}, naturally support faceted geometries. At the same time, with the progress in parallel computation, the simulation of scattering from discretized curved bodies using ray-based approaches is becoming relatively more attractive. The above-mentioned considerations led us to the choice of addressing planar-facet discretization of curved surfaces in the present work.

\rev{When curved objects are represented by planar facets, accurate modeling of diffraction becomes increasingly important. 
The Uniform Theory of Diffraction (UTD) provides an efficient framework for this purpose and remains the standard diffraction model in high-frequency ray-based simulations \cite{kouyoumjian1974uniform}.
}
However, classical UTD assumes electrically large edges and may lose accuracy for the small facets that naturally arise in discretized curved meshes.
Although advanced formulations such as vertex diffraction \cite{albani2009utd} - that can complement UTD edge-diffraction, extending its validity to the case of small edges - and higher-order diffraction \cite{albani2005uniform} are available in the literature, they are rarely integrated into practical propagation-oriented RT frameworks. 
In our previous work \cite{ziganshin2025ray}, we demonstrated that vertex diffraction can significantly improve the accuracy of scattering predictions for discretized objects. 
The present work further investigates and systematically validates this approach in the context of curved geometries.

A particularly challenging aspect is the modeling of the shadow region behind vehicles (forward scattering). Accurate prediction of the total field in the shadow often requires the inclusion of complex diffraction mechanisms, such as higher-order diffraction \cite{albani2005uniform} or creeping-wave diffraction \cite{pathak2013uniform, huang2023research}. 
Although such effects are rarely considered in ray-tracing, the present work investigates the possibility of modeling blockage from metallic bodies using double-order diffraction under a proper discretization strategy.

\rev{
Existing literature primarily addresses either accurate scattering prediction using full-wave or PO-based methods \cite{davidson2010computational, weinmann2006ray, 18706}, direct curved-surface modeling using parametric representations \cite{sefi2003pay}, or scalable propagation simulation using RT \cite{7152831, sadovaya2020ray}. However, the practical modeling of scattering and blockage from discretized curved vehicular bodies within modern RT frameworks remains insufficiently investigated.
} 
In this work, we address the challenges of modeling scattering from discretized curved bodies using standard ray-tracing tools properly extended to account for the above-mentioned diffraction interactions.
To the best of the Authors' knowledge, this is among the first studies where the foregoing approach is attempted and applied to the computation of vehicles' bistatic, near-field reflectivity. 

The main contributions of the work are the following:

\begin{itemize}
    \item Extension of a practical RT framework with advanced diffraction mechanisms, including vertex diffraction and higher-order diffraction interactions. \rev{In particular, a heuristic combination of edge and vertex diffraction is proposed to better reproduce the scattered field in the shadow region.}
    
    \item Systematic investigation of scattering and blockage modeling from discretized curved bodies, including a discretization strategy relating facet size, curvature radius, and wavelength.
    
    \item Validation against analytical solutions, full-wave EM simulations, \rev{and controlled vehicular measurements.}
\end{itemize}

The paper is organized as follows. The ray-based simulation method is described, together with the reference electromagnetic models used for validation, in Section II. The discretization strategy and its impact on the accuracy of results are discussed in Section III. In Section IV, results are presented, and the proposed approach is validated against reference models. Finally, open challenges are discussed, and conclusions are drawn in Section V.
\section{The Simulation Framework}

This section outlines the simulation setup and tools.
First, the reference electromagnetic (EM) models and analytical solutions for certain geometric primitives are introduced. 
Then, the ray-tracing framework and its extensions are presented.

\subsection{Reference Models and Test Objects}

To evaluate the proposed ray-tracing modeling approach, both analytical solutions, when available, and full-wave electromagnetic simulations are used, together with a set of canonical and realistic test geometries.

Full-wave electromagnetic simulations are performed using FEKO \cite{feko} with the multilevel fast multipole method (MLFMM) solver. 
All objects considered in this work are assumed to be perfectly electrically conducting (PEC), as vehicle bodies can often be approximated by PEC boundary conditions.

Two canonical geometries are considered to analyze discretization effects on scattering behavior: a smooth circular cylinder and a smooth sphere. 
The analytical solutions \cite{balanis2012advanced} are known for such canonical objects. 
These solutions assume plane-wave incidence, while observation points are located at a finite distance (in the near-field). 
Although the canonical validation cases assume far-field excitation, the developed RT framework naturally supports near-field transmitter configurations, which are relevant for vehicular and ISAC scenarios.

In addition to the canonical objects, realistic vehicle models are considered to evaluate the proposed approach in a more practical near-field scenario. 
\rev{Two levels of geometric complexity are investigated: a detailed vehicle mesh for backscattering validation against measurements and a simplified low-polygon representation for shadow region analysis.}
Both models represent only the external metallic body of the vehicle, while fine details such as mirrors, wheels, and other non-PEC components, as well as windows, are omitted to focus on the dominant scattering mechanisms.

\subsection{Extended Ray-tracing Framework}
Sionna-RT (v0.19) \cite{hoydis2023sionna} was used as a basic RT framework, which is open-source, parallelizable, differentiable, and provides efficient path tracing and electromagnetic field computation on triangulized geometries.
The standard RT tool has been customized with several diffraction methods to achieve better accuracy for scattering computation from discretized curved objects. The modified implementation is publicly available at \cite{sionna_rt_extension}.

In our preliminary work \cite{ziganshin2025ray}, it was shown that \textbf{vertex diffraction} is crucial to increase the accuracy of the simulation. Vertex diffraction occurs at points where several edges intersect. It complements UTD edge diffraction, as the latter assumes infinitely long edges, while edge+vertex diffraction provides a continuous field for finite edges \cite{albani2009utd}. 
Approximation of a double-curvature surface by a discretized mesh results in edges of electrically small length, for which regular edge UTD cannot yield an accurate field, as contribution from the endpoints is ignored - this is exactly what vertex diffraction is designed to fix. More details of the vertex diffraction formulation are presented in Appendix A.

Accurate modeling of the shadow region is essential in bistatic and near-field vehicular scenarios, where vehicles obstruct the direct path. 
Reliable prediction of the total field in such regions is therefore required for accurate channel modeling and sensing performance analysis.

We distinguish two ray-based methods for modeling the field in the shadow region: creeping-wave diffraction \cite{pathak2013uniform, huang2023research} and higher-order diffraction. 
The former assumes propagation along a smooth surface and is not considered in this work due to its high computational complexity.
\textbf{Higher-order diffraction} operates directly on the faceted geometries.
By chaining diffraction mechanisms, rays can propagate into shadowed regions. Although this approach is physically approximate, it is computationally compatible with standard RT frameworks and, as shown in Section IV, can provide good accuracy even for less-detailed meshes.

However, higher-order diffraction creates two main challenges. 
First, the number of propagation paths increases drastically even for the second-order diffraction. 
Second, the electromagnetic formulation of diffraction becomes increasingly complicated. While analytical solutions exist for double-edge diffraction \cite{albani2005uniform} and even for triple-edge diffraction \cite{carluccio2012utd}, the resulting expressions are cumbersome and difficult to implement in practice.
Moreover, configurations involving mixed mechanisms, like edge and vertex combinations, are not explicitly addressed in the existing literature. 
Such interactions frequently occur in discretized meshes where diffraction points may lie close to edge vertices. 

Our approach for higher-order diffraction includes several 2nd order propagation phenomena.

1) \textbf{Double-edge diffraction (EE)}. The electromagnetic formulation of double-edge diffraction is available for the arbitrary configuration of edges \cite{albani2005uniform}. The formulation of double-edge diffraction is presented in Appendix A as well. 

The computational complexity of finding diffraction points depends strongly on the geometric configuration. Two configurations can be distinguished:

a) \textbf{coplanar (including parallel) edges} – a closed-form solution exists for determining diffraction points. This configuration is typical for discretized convex bodies, where adjacent edges lie on the same planar facet. In this work, double-edge diffraction is restricted to this case.

b) \textbf{non-coplanar edges} - no closed-form solution exists for determining diffraction points; the problem should be solved via numerical optimization, significantly increasing computational complexity \cite{carluccio2008efficient}. 
This feature is not critical in the scope of the current work, as objects are primarily assumed to be convex, so this case is not considered.

By restricting EE diffraction to coplanar adjacent edges, computational complexity remains manageable while still improving the modeling of the shadow region in convex geometries.

\textbf{2) Edge–Vertex and Vertex–Edge Diffraction (EV, VE)}
In addition to EE diffraction, cascaded edge–vertex (EV) and vertex–edge (VE) diffraction paths are implemented, as they are also essential for accurate shadow modeling.
Since a rigorous formulation of EV/VE diffraction is not available in the literature, a heuristic cascading approach is adopted.
More details and limitations on this approach are discussed in Appendix A.
\rev{Therefore, the objective is not to derive a rigorous higher-order diffraction formulation but to evaluate whether such practical approximations can improve RT prediction.}

\begin{figure}[t]
\centering

\subfloat[The EE path passes close to the vertex, leading to a significant EV contribution \label{fig:ev_1}]{\includegraphics[width=0.65\linewidth]{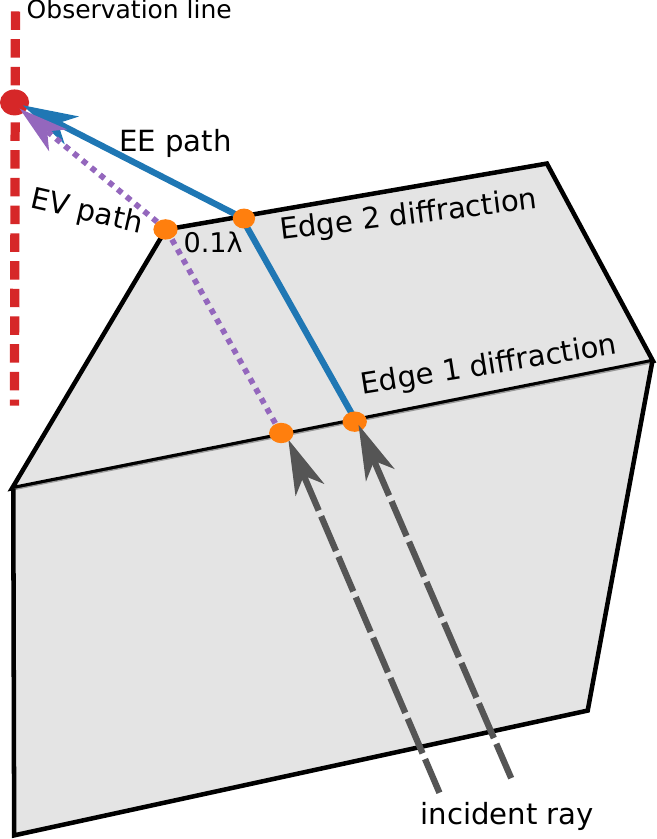}}

\subfloat[Total electric field magnitude: Extended RT vs MLFMM \label{fig:ev_2}]{\includegraphics[width=0.95\linewidth]{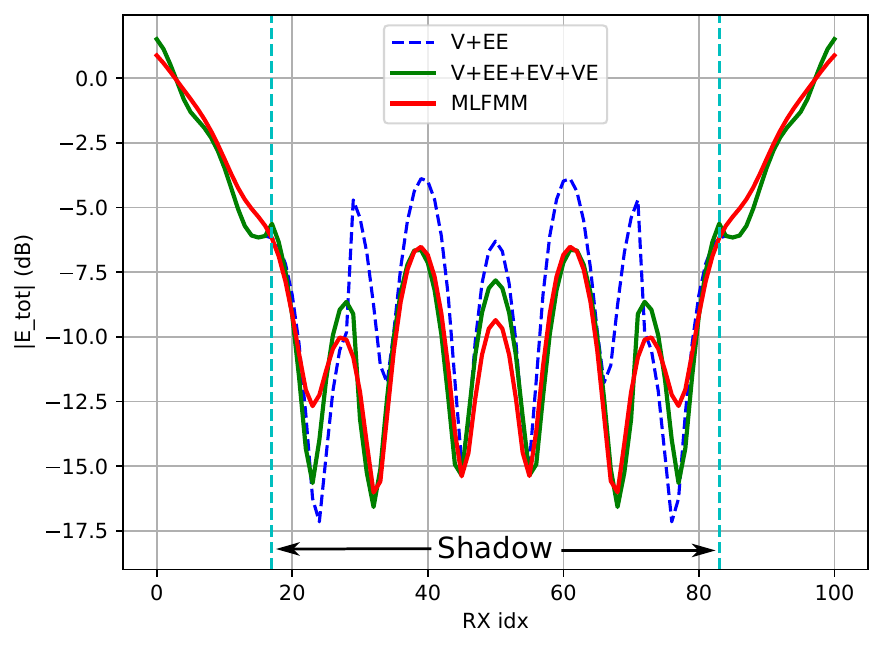}}

\caption{Example used to illustrate the importance of edge–vertex diffraction}
\label{fig:ev_path}
\end{figure}

To illustrate the importance of edge–vertex diffraction, a simple geometric configuration (\cref{fig:ev_1}) was constructed in which an incident ray interacts sequentially with two edges.
The interaction point at edge 2 is at $0.1 \lambda$ distance from the vertex, therefore the field is strongly influenced by the vertex contribution. 
The resulting field is evaluated along a line of observation points placed behind the object.

The results are in \cref{fig:ev_2}. When only vertex and double-edge diffraction are considered (V+EE), the predicted field exhibits noticeable discrepancies with respect to the full-wave reference solution. In particular, the absence of EV and VE mechanisms leads to incorrect behavior in the shadow region.
When edge–vertex and vertex–edge contributions are included (V+EE+EV+VE), the agreement with the MLFMM solution improves significantly.

This example illustrates that vertex-related interactions are essential for accurate modeling of the shadow region of discretized objects.
However, certain limitations of the double diffraction formulation remain, which are also highlighted in Appendix A.

\subsection{Discretization strategy and tradeoffs}

In ray-tracing frameworks, curved objects are approximated by planar facets, since native support for smooth parametric surfaces (e.g., NURBS) is typically unavailable in standard RT tools such as Sionna-RT. This introduces a trade-off between geometric fidelity, electromagnetic accuracy, and computational cost.

Under-discretization yields a poor geometric approximation, degrading the angular smoothness of the scattered field. Over-discretization increases the number of edges and diffraction paths requiring higher-order diffraction ($>2$) to reach shadow regions and increases computation time: single-edge diffraction scales as $O(N_e)$, while second-order diffraction scales as $O(N_e^2)$, rapidly becoming prohibitive. Moreover, electrically small edges violate basic UTD assumptions, producing inaccurate diffraction contributions.

In vehicular scenarios, meshes are typically obtained from LiDAR scans, CAD models, or public repositories, and require post-processing prior to electromagnetic use (removal of small details, correction of non-manifold edges, and intersecting facets). A key step in this work is curvature-based remeshing~\cite{levy2013variational}.

The discretization criteria stem from the geometric approximation error of representing a smooth surface with planar facets, which results in the electromagnetic phase error relative to the wavelength. To formalize the problem, we introduce: $R$ (local radius of curvature), $E$ (characteristic facet edge length), $s$ (maximum geometric deviation), and $\lambda$ (wavelength).

Approximating a curved surface with facets introduces geometric deviation. 
From an electromagnetic perspective, this geometric deviation produces a phase error $\Delta \phi \sim 2\pi s / \lambda$, which should remain a small fraction of $2\pi$ to maintain a reasonable accuracy.
\rev{Maintaining a sufficiently small phase error is particularly important for deterministic scattering prediction. However, the acceptable error levels should depend on the application; for instance, statistical channel modeling applications may tolerate larger phase deviations.}

\rev{Therefore, the physically relevant quantity is the normalized geometric deviation $s/\lambda$, which directly controls the discretization-induced phase error. 
In practice, however, $s$ is generally difficult to estimate for arbitrary meshes because the underlying smooth surface is often unknown and only the discretized representation is available.}

Assuming $E \ll R$, the deviation can be estimated from the geometric distance between a small arc and its corresponding chord of length $E$.
It can be easily shown that the deviation scales as $s \sim E^2 / R$.
This leads to the parameter $E^2 / (R \lambda)$, which \rev{serves as a practical mesh-based proxy for $s/\lambda$} and governs the accuracy of the discretized representation.

\rev{The parameter $E^2/(R\lambda)$ therefore provides a physically motivated measure of discretization quality that can be evaluated directly from the mesh geometry.
The optimum operating range of this parameter depends not only on geometric accuracy but also on the desired application, considered propagation mechanisms, and limitations of the ray-based model.
Consequently, the practical operating range needs to be established empirically through comparison against analytical and full-wave reference solutions.}

In addition to the curvature-related metric, the facet size should be of several wavelengths in order to satisfy the UTD assumptions, typically $E > 1.5 \lambda$. 
Even though the UTD theory with vertex diffraction is always stable and can produce consistent results, electrically small edges may degrade accuracy and may introduce instability in heuristic diffraction methods.

\rev{The validity and limitations of the proposed discretization metric are investigated in the following section using canonical PEC geometries.}

\section{Validation and Numerical Results}

This section evaluates the proposed method on both canonical and realistic geometries, using analytical and full-wave EM solutions as references. 
Two distinct error sources are present: discretization and RT approximation error. Comparisons with analytical solutions are subject to both, since closed-form solutions exist only for smooth bodies.
Comparisons with full-wave MLFMM simulations isolate the RT modeling error since both methods can operate on the same discretized geometry.

\begin{figure}[t]
    \centering
     \includegraphics[width=0.75\linewidth]{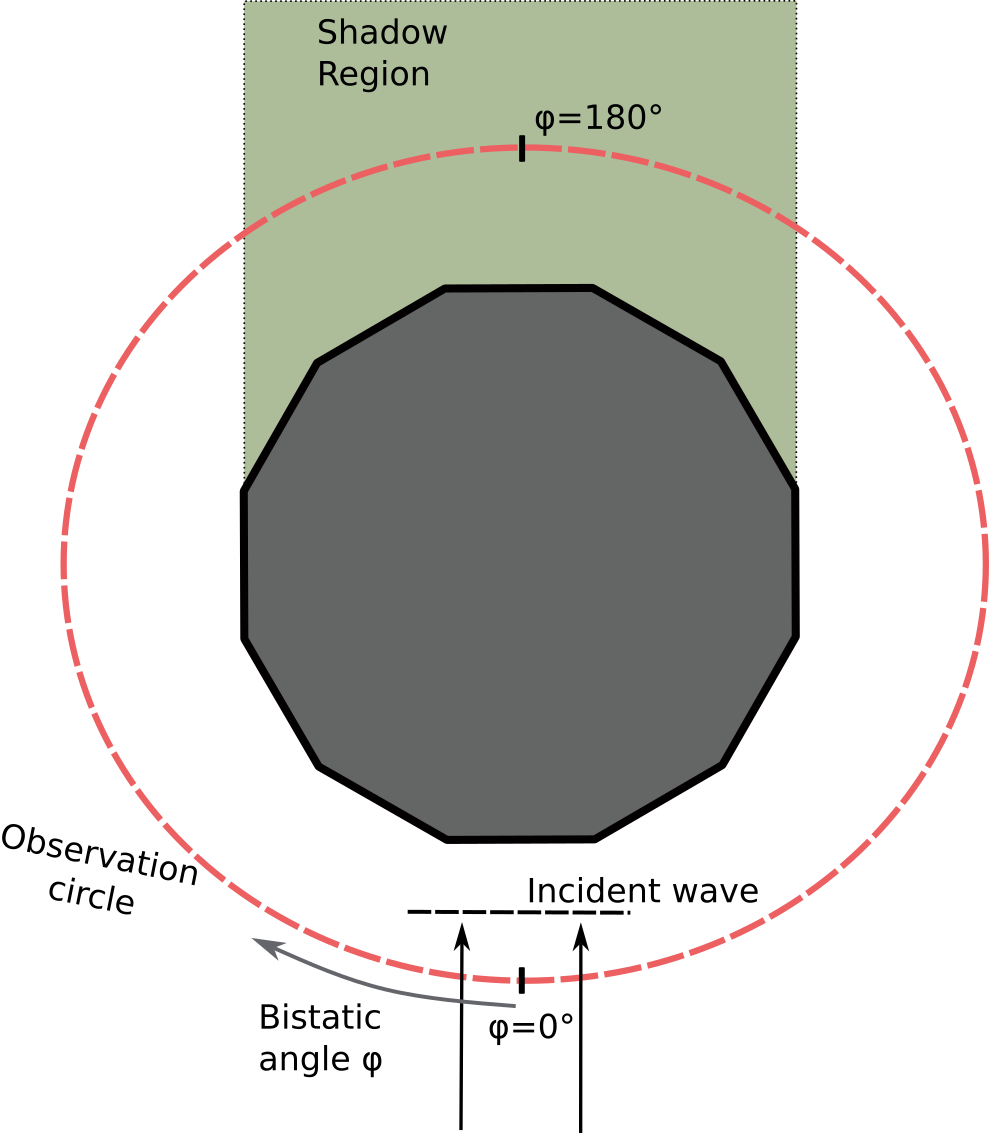}
    \caption{General simulation geometry used for scattering evaluation.}
    \label{fig:scheme}
\end{figure}

The general simulation setup is illustrated in \cref{fig:scheme}. The object is illuminated under either plane-wave or near-field conditions, while observation points are placed on a circular trajectory around the object. The bistatic angle is defined between the incident and observation directions. The region opposite to the incident direction corresponds to the shadow region. $0^\circ$ corresponds to when the source and observation are aligned in the same direction, and $180^\circ$ when they are on opposite sides.
In the following result figures, the boundaries of the shadow region are indicated by dashed cyan lines.

Both the scattered and total electric fields are analyzed: the former is more informative in the backscattering region, where the incident field would overshadow the scattered field, while the latter is better suited for the shadow region. The validation proceeds from canonical objects with known analytical solutions toward a realistic vehicular geometry. All objects are assumed to be PEC; the simulated frequency is 2 GHz; antenna patterns are assumed for simplicity omnidirectional.

\subsection{Canonical Geometries}

\begin{figure}[t]
    \centering
    \includegraphics[width=0.96\linewidth]{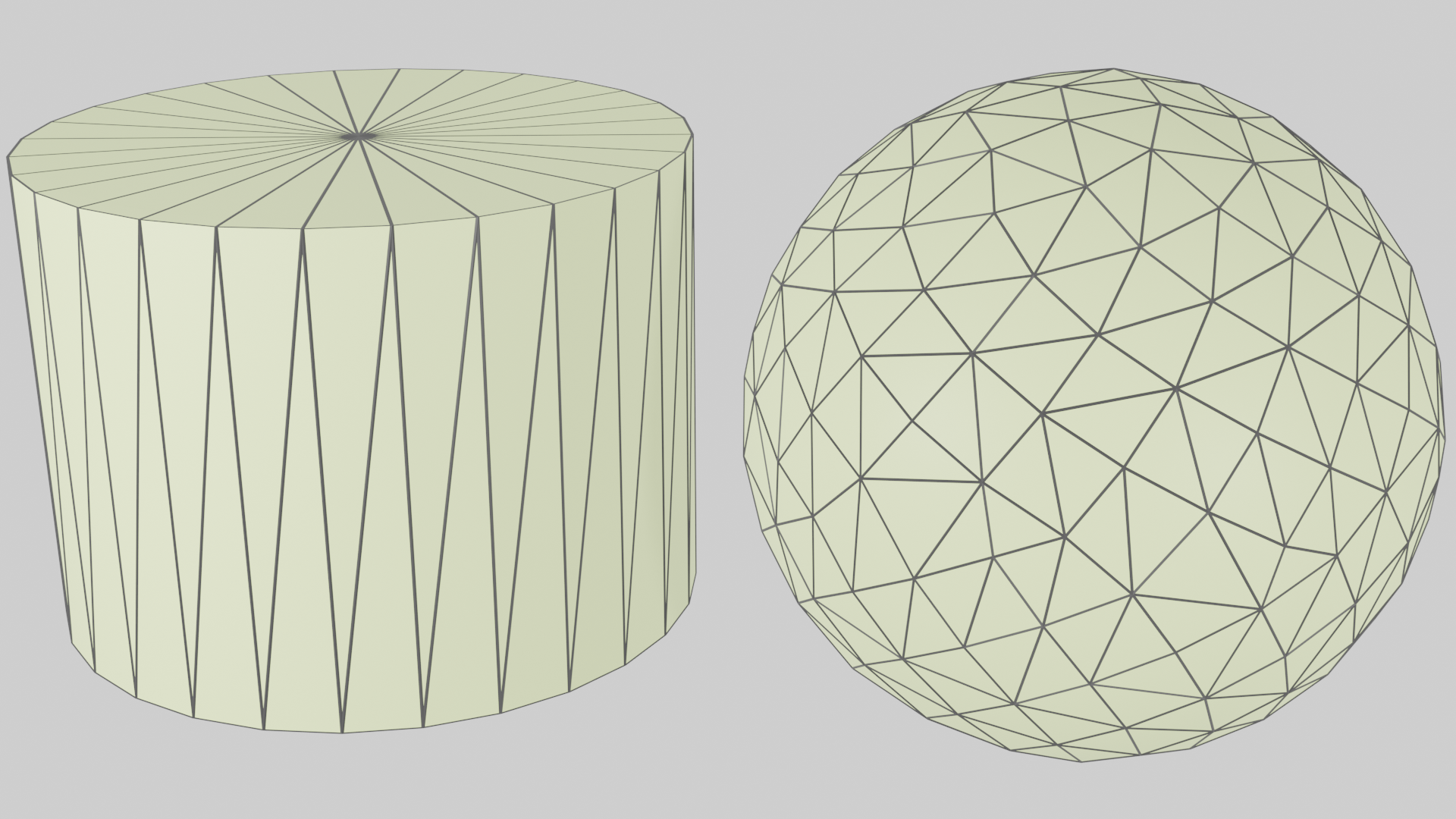}
    \caption{Examples of discretized canonical geometries used for validation: a circular cylinder (left) and a sphere (right).}
    \label{fig:can_objects}
\end{figure}

\begin{figure*}[t]
\centering
\subfloat[Scattered field, cylinder \label{fig:cyl_sc}]{\includegraphics[width=0.45\textwidth]{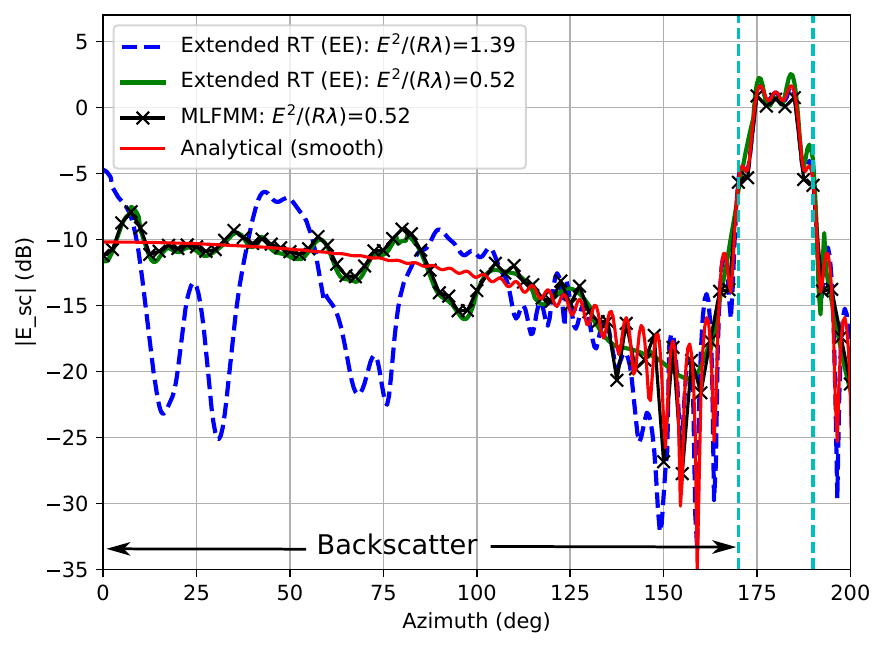}}
\hfill
\subfloat[Total field, cylinder: $E^2 / (R \lambda) = 1.39$ \label{fig:cyl_tot}]{\includegraphics[width=0.45\textwidth]{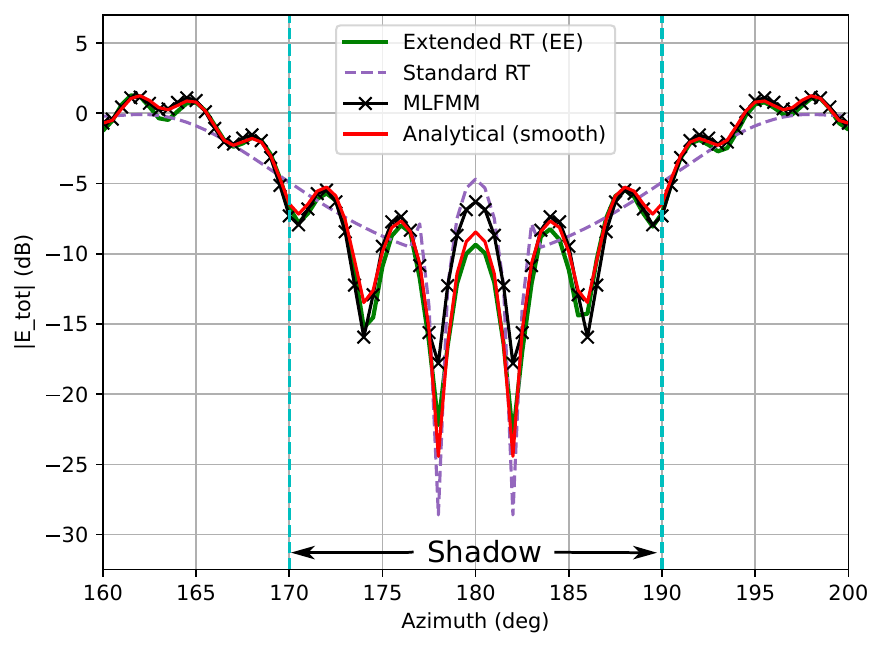}}
\hfill
\subfloat[Scattered field, sphere: $E^2 / (R \lambda) = 0.61$ \label{fig:sphere_sc}]{\includegraphics[width=0.45\textwidth]{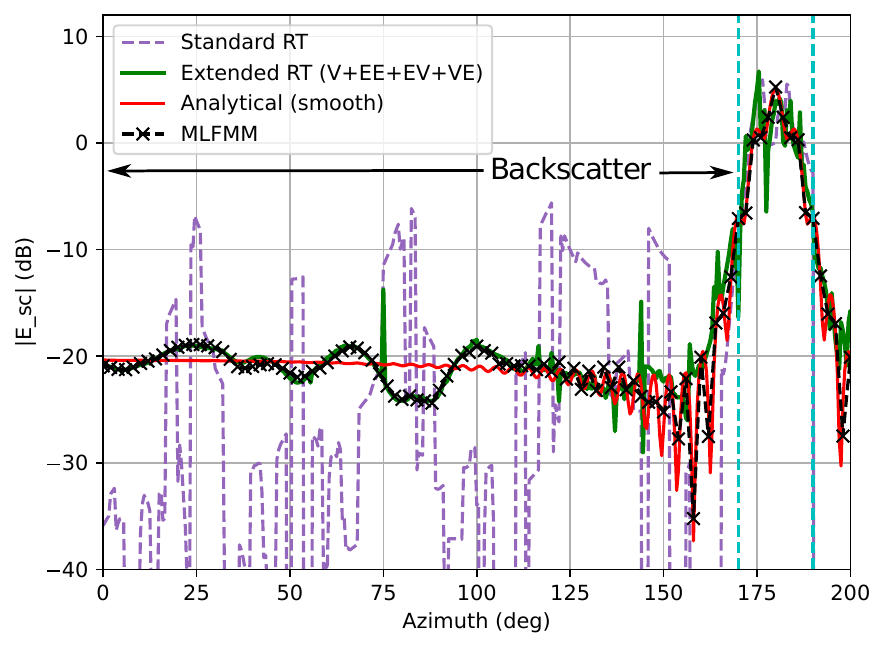}}
\hfill
\subfloat[Total field, sphere: $E^2 / (R \lambda) = 1.34$ \label{fig:sphere_tot}]{\includegraphics[width=0.45\textwidth]{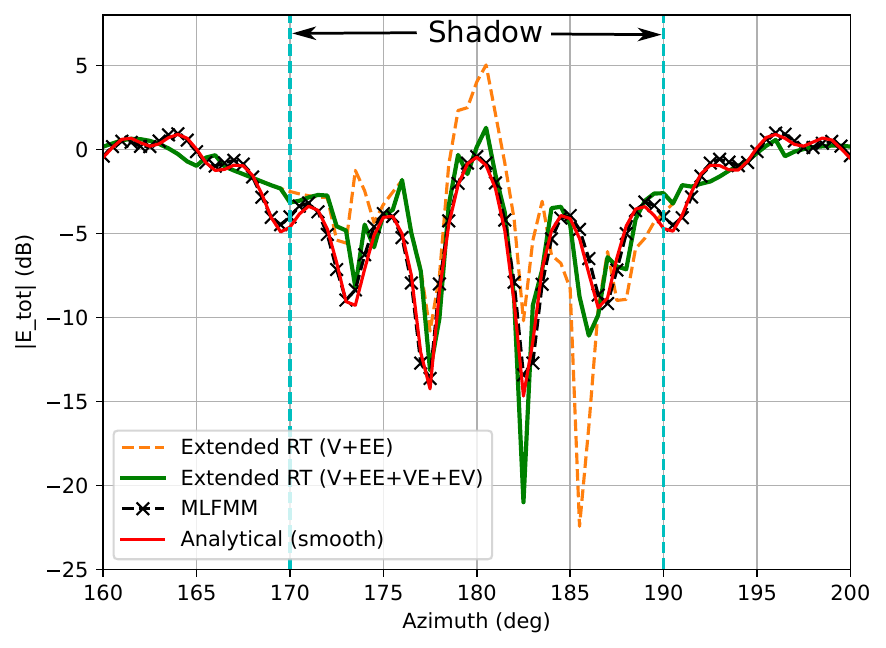}}
\caption{Electric field magnitude, HH polarization: RT vs. MLFMM vs. the smooth-object analytical solution}
\label{fig:fields_can_objects}
\end{figure*}

To evaluate the proposed framework, canonical geometries with known analytical solutions are considered. 
\rev{The validation addresses three questions:
\begin{itemize}
    \item How does the discretization level affect scattering prediction?
    \item What is the impact of the proposed diffraction extensions?
    \item Can a practical discretization guideline be identified?
\end{itemize}
}

A circular cylinder and a sphere of radius $7 \lambda$ are used as representative examples of smoothly curved bodies that are approximated by planar facets. 
Both objects are discretized into triangular surface elements, as exemplified in \cref{fig:can_objects}.
Discretized cylinders are assumed infinitely long in the analytical solution, while in the simulations, they are chosen long enough so that diffraction over the top and bottom parts does not significantly affect the results.
The setup follows the general configuration in \cref{fig:scheme} with plane-wave illumination.
Observation points are located on the circle around the object at a distance of $40.0\lambda$ with angular separation $0.5^\circ$, yielding 720 points.
\rev{The agreement between RT and reference solutions is quantified using the mean absolute error (MAE) between the angular field responses in dB.}

For comparison, a baseline ray-tracing configuration, referred to as \emph{Standard RT}, is also considered. 
Standard RT corresponds to the default Sionna-RT implementation with specular reflections and single-bounce edge diffraction, without the proposed vertex and higher-order diffraction extensions.

\subsubsection{Cylinders}

The comparison of the \textbf{scattered} field predicted by ray-tracing at two discretization levels against the analytical solution under HH polarization (VV polarization behaves similarly) is shown in \cref{fig:cyl_sc}.
Due to the symmetry of the object, the region of angles $0^\circ - 200^\circ$ is considered.
RT with coarse mesh (dashed blue curve) exhibits noticeable deviations from the analytical solution.
Refining the discretization (green) significantly improves agreement with the analytical solution.
Comparison to MLFMM shows good agreement in the backscattering region and worsening close to the shadow region, where multiple bounces are required to accurately describe the field.

The \textbf{shadow} region for quite coarse discretization is shown in \cref{fig:cyl_tot}. 
The standard RT approach (dashed purple) falls short in reproducing the field in the shadow accurately. 
The inclusion of double-edge diffraction (green curve) improves the prediction from 2.3 to 0.8 dB of MAE and reproduces the behaviour observed in the analytical solution.
Meanwhile, the MLFMM curve is also similar to the analytical solution.

\subsubsection{Sphere}

The sphere represents a more challenging benchmark due to its double curvature. 
In this case, the discretization generates a larger number of finite edges, increasing the importance of vertex-related diffraction mechanisms.

The \textbf{scattered} field for the relatively fine discretized sphere is presented in \cref{fig:sphere_sc}.
The standard RT implementation (dashed purple) exhibits strong discontinuities and noticeable deviations from the analytical reference.
The inclusion of vertex diffraction (green) significantly improves field continuity and leads to substantially better agreement with the analytical as well as MLFMM references.

Several diffraction configurations in the \textbf{shadow} region are compared in \rev{\cref{fig:sphere_tot}.}
RT with EE diffraction (orange) enables propagation into the shadowed area.
The further inclusion of EV and VE interactions (green) provides additional refinement of the predicted field, and MAE improves from 3.5 to 2.2 dB.

\rev{
To quantify the contribution of the individual diffraction mechanisms, \Cref{tab:diffraction_improvement} summarizes the average MAE in the sphere shadow region for different RT configurations.
Only coarser discretizations with $E^2/(R\lambda)>0.6$ are considered, as these cases are most representative for practical shadow-region prediction and exhibit the strongest influence of higher-order diffraction mechanisms.
For observation angles where no valid propagation path is predicted by the considered RT configuration, a penalty error of 20 dB is assigned.
}

\begin{table}[t]
\caption{Average MAE in the sphere shadow region for different RT configurations and polarizations,  with discretizations $E^2/(R\lambda)>0.6$.}
\label{tab:diffraction_improvement}
\centering
\begin{tabular}{lccc}
\hline
Polarization & V & V+EE & V+EE+EV+VE \\
\hline
HH & 7.3 dB & 4.0 dB & 3.0 dB \\
VV & 8.3 dB & 3.2 dB & 2.7 dB \\
\hline
\end{tabular}
\end{table}

\subsubsection{Discretization Sensitivity}

\begin{figure}[t]
    \centering
    \includegraphics[width=0.95\linewidth]{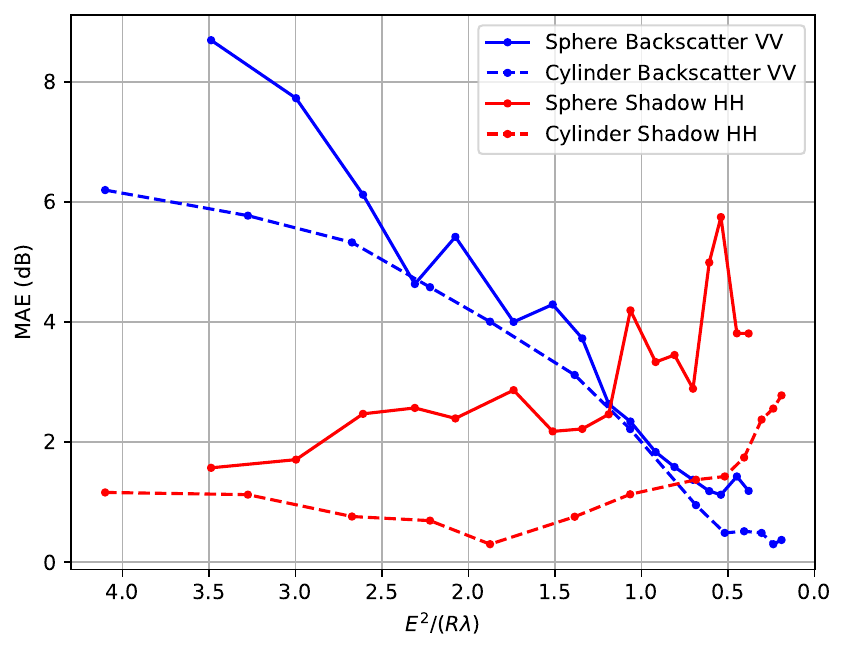}
    \caption{MAE relative to the analytical solution as a function of $E^2/(R\lambda)$ for the considered canonical geometries and propagation regions.}
    \label{fig:discr_sens}
\end{figure}

\rev{To investigate the generality of the proposed discretization strategy, the MAE between ray-tracing and analytical solutions is evaluated for multiple discretization levels.
The error as a function of the parameter $E^2 / (R \lambda)$ is summarized in \cref{fig:discr_sens}.
Lower values of $E^2 / (R \lambda)$ correspond to finer discretization.
Several observations can be made.}

\rev{First, in the backscattering region, accuracy generally improves as the discretization becomes finer. However, the improvement becomes progressively smaller once $E^2 / (R \lambda)$ approaches approximately 0.5. Further refinement provides only marginal accuracy improvement but increases the computational burden.}

\rev{Second, in the shadow region, excessively fine discretization may become counterproductive. Even relatively coarse meshes reproduce the overall shadow behavior reasonably well.
This behavior can be physically explained by the fact that the geometric discretization error does not directly translate into an equivalent propagation path error. 
Paths propagating tangentially along the object surface experience significantly smaller path length variations, making shadow-region prediction less sensitive to discretization accuracy.}
\rev{Moreover, since the present implementation is limited to second-order diffraction, higher diffraction orders that may be required for finer discretization levels cannot be represented.}

Finally, when a single mesh simultaneously has to be used for both backscattering and shadow region prediction, the interval $E^2 / (R \lambda) \approx 0.6 - 0.9$ provides a practical compromise between accuracy in both regions and computational complexity.

\subsection{Vehicular Scattering Validation}

\begin{figure*}[t]
\centering

\subfloat[BIRA setup with the car \label{fig:bira_car}]{\includegraphics[width=0.31\textwidth]{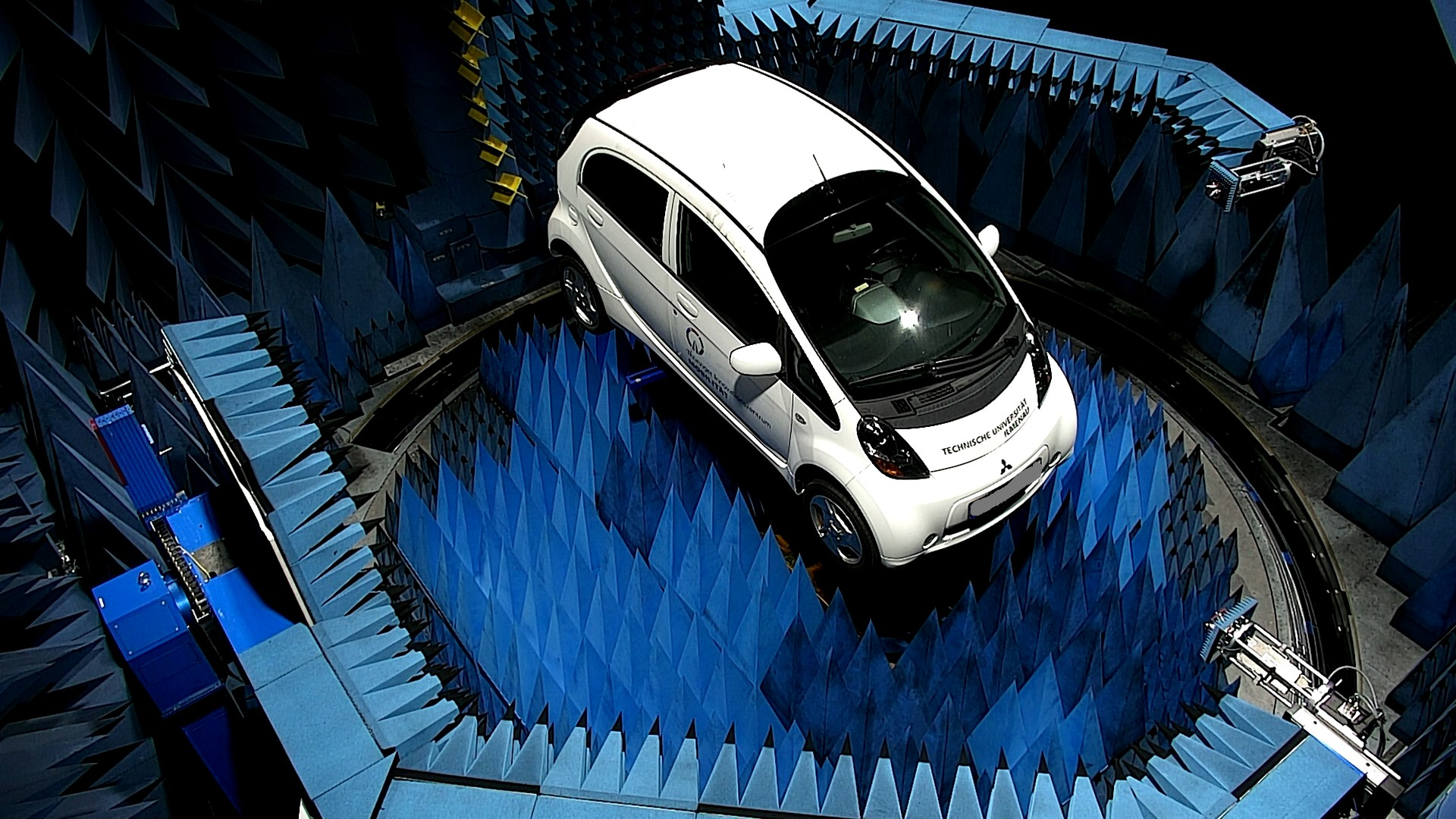}}
\hfill
\subfloat[Detailed mesh (1496 facets) \label{fig:detailed_mesh}]{\includegraphics[width=0.31\textwidth]{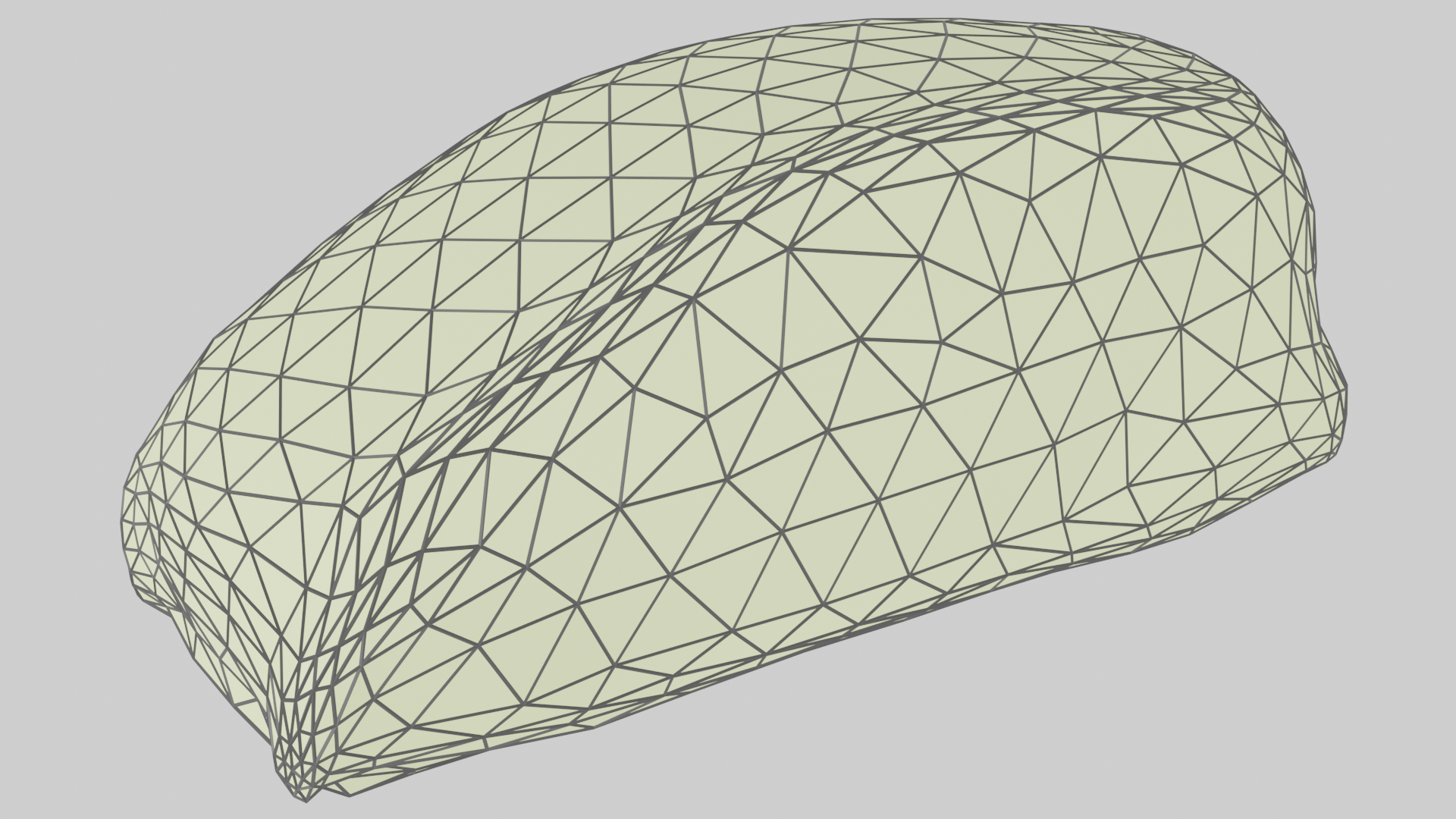}}
\hfill
\subfloat[Low-polygonal mesh (220 facets) \label{fig:low_poly_mesh}]{\includegraphics[width=0.31\textwidth]{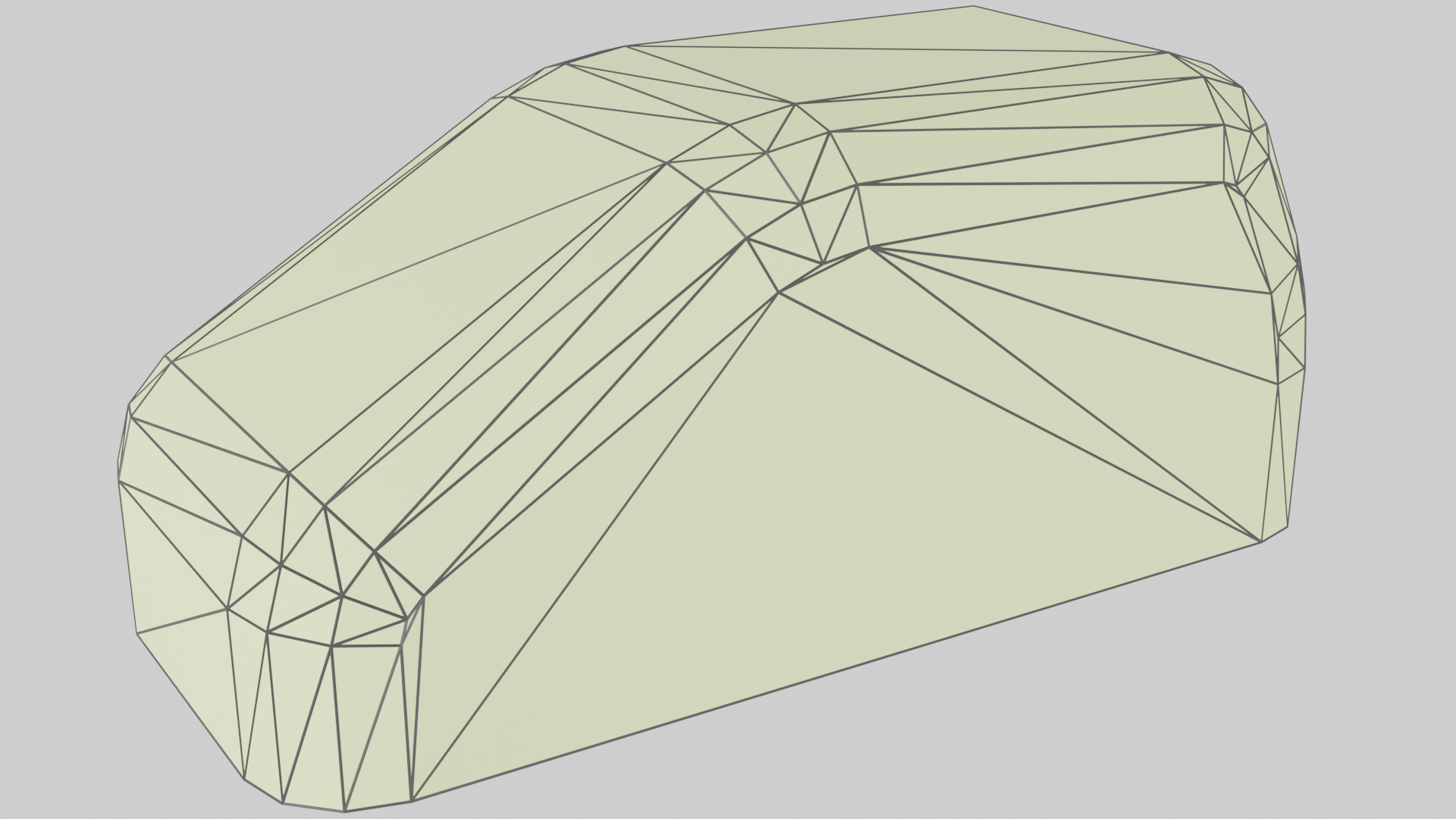}}

\caption{Vehicle and mesh representations used for the validation study.}
\label{fig:car_mesh}
\end{figure*}

\begin{figure*}[t]
\centering
\subfloat[Measurement: 3 GHz, HH polarization \label{fig:comp_back}]
{\includegraphics[width=0.45\textwidth]{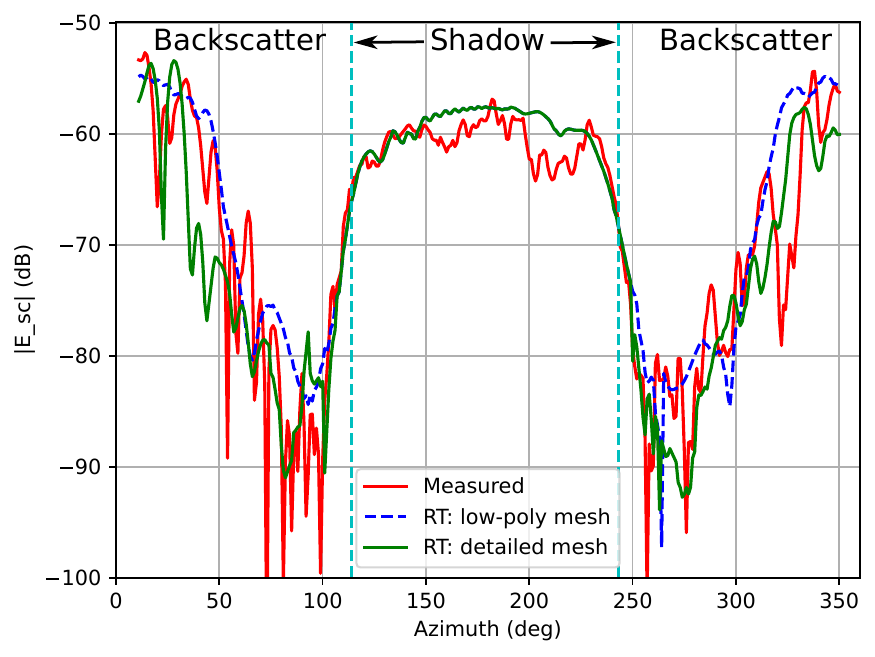}}
\hfill
\subfloat[MLFMM: low-poly mesh, 2 GHz, VV polarization \label{fig:comp_total}]
{\includegraphics[width=0.45\textwidth]{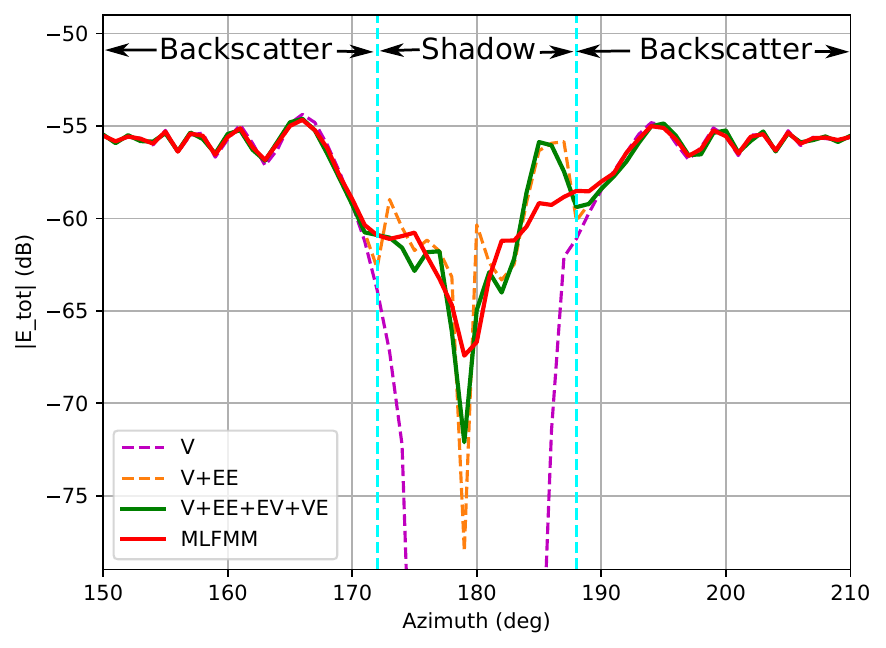}}

\caption{Vehicle-case validation of the extended RT framework.}
\label{fig:low_poly_car_two}
\end{figure*}

The previous section demonstrated the validity of the proposed framework using canonical PEC geometries. 
In this section, a more realistic vehicular scenario is considered to assess the practical relevance of the method.

Based on the previous analysis, different mesh representations are expected to be beneficial for different propagation mechanisms. 
\rev{Backscattering prediction benefits from a finer geometric representation, whereas shadow region prediction can often be achieved using considerably coarser discretizations. 
Therefore, two complementary validation scenarios are considered:}

\rev{
\begin{itemize}
\item comparison against measured vehicle backscattering using both detailed and low-polygon vehicle meshes
\item comparison against MLFMM simulations in the shadow region using a low-polygon vehicle model
\end{itemize}
}

\rev{
The measurement comparison is restricted to the backscattering region, where the dominant response from the external vehicle geometry is expected to be reasonably approximated using a PEC model. 
In contrast, shadow region prediction is strongly influenced by transmission through windows, dielectric vehicle components, and additional scattering mechanisms that are not represented in the actual model. 
Therefore, the shadow region validation is performed using controlled numerical comparisons against MLFMM, and the incorporation of realistic material properties is left for future work.
}

\rev{
The measurement setup together with the vehicle meshes used in this study is shown in \cref{fig:car_mesh}.
The detailed mesh with 1496 facets (\ref{fig:detailed_mesh}) is employed for validation against measurement in backscattering.
Although the mesh is detailed, it was simplified by removing several complex parts, such as wheels and other non-PEC elements.
After that, it was curvature-adaptively remeshed \cite{levy2013variational}.
The parameter $E^2 / (R \lambda)$ at 3 GHz is estimated as around $0.4-0.6$.
The low-polygon representation (\ref{fig:low_poly_mesh}) consists of 220 facets and is used for the shadow region analysis.
}

\subsubsection{Comparison Against Bistatic Measurements}
\rev{The measurements were performed using the BIRA bistatic reflectivity setup \cite{andrich2026bira} with frequency range 2-18 GHz. 
A small vehicle, Mitsubishi i-MiEV, depicted in \cref{fig:bira_car}, was illuminated from the side, and the field was recorded over a wide angular range with $1^\circ$ resolution.
The distance from the car to the antennas is  $3.0 \, \mathrm{m}$; height from the car bottom is  $0.75 \, \mathrm{m}$.
}

\rev{
Post-processing of the data includes background subtraction, time-delay gating using car dimensions, and LOS calibration \cite{andrich2025wideband}.
Due to calibration limitations, the lower and upper band edges exhibit increased uncertainty. 
Therefore, a representative frequency of 3~GHz is selected for the comparison.
MLFMM results are not included in this comparison because the available full-wave setup does not reproduce the measured antenna patterns, which would make a direct comparison misleading.
}

\rev{
The comparison in \cref{fig:comp_back} shows the measured scattered field together with the predictions obtained using the detailed and low-polygon vehicle meshes.
Both meshes reproduce the dominant scattering behaviour and trends observed in the measurements. 
However, the detailed mesh provides a closer agreement with the measured response and more realistic angular variations.
}

\rev{
The comparison should be interpreted as a qualitative validation rather than an attempt to exactly reproduce the measurements. 
The present simulations assume PEC surfaces and neglect dielectric vehicle components. 
Additional discrepancies may also arise from measurement uncertainty, background subtraction, antenna pattern inaccuracies, and positioning errors.
}

\subsubsection{Shadow Region Validation Against MLFMM}
To further evaluate the shadow region prediction, the low-polygon vehicle model is compared against the MLFMM simulation.
The object is illuminated from the side at a distance of $20 \, \mathrm{m}$, while observation points are located on a circle at a radius of $20 \, \mathrm{m}$ with the angular separation of $1^\circ$; height of the source and observation points is $0.6 \, \mathrm{m}$.

The total field at 2~GHz obtained using the proposed RT framework and the corresponding MLFMM reference solution is shown in \cref{fig:comp_total}.
The single-bounce RT (dashed purple) cannot reach the deep shadow. 
Inclusion of EE (dashed orange) improves the shadow region results drastically. 
Inclusion of EV and VE combinations (green) helps further improve the result, MAE from 2.6 to 1.6 dB. 

\rev{
While the agreement is promising, realistic vehicular shadowing scenarios are expected to exhibit additional complexities associated with material-dependent transmission and scattering effects that are not considered in the present study.
}

\subsection{Computational Performance}
Recent advances in multi-core processors and graphical processing units have significantly accelerated ray-based simulation frameworks. 
Tools such as Sionna enable efficient path tracing and electromagnetic field evaluation on discretized geometries. 

\rev{
All simulations were performed on a CPU platform. 
Since the considered scenarios involve a relatively small number of computational tasks, they do not significantly benefit from GPU acceleration.
Also, all observation locations are processed simultaneously (in parallel) within a single RT simulation.
The computational performance was evaluated on a dual-socket AMD EPYC 7343 CPU platform with 32 physical cores.
Reported runtimes correspond to the average of 5 independent runs.
}

Although a detailed comparison with full-wave electromagnetic solvers is beyond the scope of this work, it should be emphasized that MLFMM simulations are considerably more computationally demanding. 
For instance, each of the MLFMM simulations in this study took several hours.
Instead, we evaluate the relative computational cost of different ray-tracing configurations depending on the propagation mechanisms included in the simulation. 
In particular, we analyze the impact of vertex diffraction and second-order diffraction mechanisms on the overall simulation time. 

\rev{
The considered scenarios include two spherical cases with 720 angular points and two vehicle meshes with 360 angular points. 
The two sphere cases correspond to the backscattering and shadow analyses: (i) a finer discretization with $E^2/(R \lambda) = 0.61$ with 346 facets, and (ii) a coarser discretization with $E^2/(R \lambda) = 1.34$ with 156 facets.
For the vehicle, the detailed (1496 facets) and low-polygon (220 facets) meshes are considered.
The simulation time results are presented in \Cref{tab:comp_time}.
}

\begin{table}[t]
\caption{Execution time (in seconds) for different geometries and RT configurations.}
\label{tab:performance}
\centering
\begin{tabular}{lcccc}
\hline
Scenario & RT & +V & +EE & +EV/VE \\
\hline
Sphere (156 facets)     & 0.3 & 1.1 & 1.8 & 4.6 \\
Sphere (346 facets)    & 0.4 & 1.8 & 3.1 & 6.8 \\
Vehicle (220 facets)     & 0.4 & 0.9 & 2.4 & 5.0 \\
Vehicle (1496 facets)     & 0.7 & 3.6 & 10.2 & 19.0 \\
\hline
\end{tabular}
\label{tab:comp_time}
\end{table}

The results demonstrate that the computational cost strongly depends on the selected mesh resolution. 
While detailed meshes are beneficial for accurate backscattering prediction, their use together with higher-order diffraction mechanisms significantly increases the computational burden with less benefit for simulation accuracy. 
\rev{For shadow region analysis, a coarser discretization may be preferable, as it reduces the number of diffraction interactions and can even improve accuracy.}

\rev{It should also be noted that the reported runtimes correspond to the current Sionna-RT v0 implementation. The recent migration of Sionna-RT to Dr.Jit-based architecture in the v2 release is expected to provide additional computational speedups.}

\section{Discussion and Conclusion}

\rev{
The representation of smooth objects with facets inherently creates a large number of vertices and finite edges.
Therefore, accurate ray-based simulation of discretized curved objects requires extending standard RT approaches with advanced mechanisms to account for diffraction from finite edges, vertices, and their higher-order interactions.
}

The results show that accurate ray-based modeling of scattering from curved bodies is governed by a coupled trade-off between geometric discretization, diffraction modeling, and computational complexity. 
\rev{The presented analysis shows that no universal mesh resolution exists for all propagation mechanisms.}
Backscattering benefits from finer geometric representations, while shadow-region prediction may be more accurately and efficiently achieved using coarser meshes. 
The proposed criterion $E^2/ (R \lambda)$ should be interpreted as a practical trade-off between geometric accuracy, diffraction modeling requirements, and computational complexity. The adoption of a dual-accuracy discretization - finer for backscattering and coarser for forward scattering - may offer a practical approach.


\rev{
Canonical validation shows that the diffraction extensions significantly improve accuracy, and mixed EV/VE diffraction mechanisms are important to achieve reasonable accuracy in shadow regions.  
However, more rigorous higher-order diffraction formulations may still be required for more accurate prediction.
The presented vehicular validation demonstrates the practical applicability of the proposed framework for realistic wireless propagation scenarios. 
The agreement with bistatic measurements in the backscattering region and with full-wave MLFMM simulations in the shadow region confirms that the proposed approach can capture the dominant scattering and blockage effects of complex vehicular objects. 
Nevertheless, the current study relies on several approximations, including the PEC surface and the neglect of transmission.
}

Therefore, the work should be considered as a computationally efficient approximation suitable for large-scale simulations.
The framework is aimed at vehicular and ISAC channel simulations where computational efficiency and physically motivated modeling are prioritized. 
Beyond channel simulation, it supports stand-alone object reflectivity analysis, with potential extensions toward statistical characterization for stochastic channel models. 

Open research challenges include:
\begin{enumerate}
\item Improved shadow-region prediction through more \rev{rigorous higher-order diffraction formulations} and additional propagation mechanisms, including transmission through dielectric objects.
\item Extension of analysis and validation to more complex, concave, and non-PEC structures.
\item Hybrid deterministic-stochastic approach for large-scale simulations.
\item Modeling of time-varying substructures (micro-Doppler) \cite{costa2025modeling}.
\item \rev{Gradient-based optimization for calibrating digital environments and object representations directly from channel measurements \cite{hoydis2024learning}}.
\end{enumerate}

\section*{ACKNOWLEDGMENT}
This work is partially supported by COST Action CA20120 through a Short-Time Scientific Mission, as well as by project 4-CAD P1, financed by Deutsche Forschungsgemeinschaft under grant GA 2062/7-1, by project BMBF 6G-ICAS4Mobility with Project No. 16KISK241, by project BIRAUM Project No. 2025FGR0020, and by project D-TRACE with Project No. 2025FGR0086.

We sincerely thank our friend, Professor Matteo Albani from the University of Siena, Italy, for his insightful suggestions on the implementation and use of UTD and vertex diffraction models.
We also thank T. Nowack, S. J. Myint, I. Varga, and M. Pourjafarian for their assistance during the measurement campaign.
\section*{APPENDIX}

This appendix summarizes the diffraction mechanisms implemented in the extended ray-tracing framework, with emphasis on their role in ensuring field continuity across shadow boundaries. 
The formulation is presented in a compact form, emphasizing the structure of the diffraction coefficients and associated transition functions.

\textbf{1) Edge diffraction}\\
In ray-based methods, geometrical optics fields undergo discontinuities at shadow boundaries, where reflected fields abruptly appear or vanish on the edges of the discretized object. 
The edge diffraction mechanism acts as a correction term that restores continuity of the field.

The UTD formulation of the diffraction coefficient \cite{kouyoumjian1974uniform} can be written in the form 
\begin{equation}
    D_E \sim \sum_{i=1}^{4} \cot \!\left(\frac{\Phi_i}{2n}\right) F\!\left(a_i\right)
    \label{eq:ordinaryUTDcoeff}
\end{equation}
where $\Phi = \pi \pm (\phi \mp \phi')$ is defined by the azimuthal angles of the incidence ($\phi'$) and diffraction ($\phi$) directions, $n$ is a parameter related to the wedge aperture, so that the wedge angle is equal to $(2-n)\pi$, $a$ defines the optical distance from the transition, and the uniform solution is governed by the transition function $F(\cdot)$, which restores continuity of the incident and reflected field across the incidence and reflection shadow boundaries, respectively.

The diffraction coefficient depends on polarization with respect to the edge-fixed incidence and diffraction planes \cite{kouyoumjian1974uniform}. 
Two canonical cases are distinguished: hard polarization, where the electric field is orthogonal to the plane of incidence/diffraction, and soft polarization, where it is parallel.

A limitation of classical edge diffraction is the assumption of infinitely long edges, which leads to inaccuracies and discontinuities when the diffraction ray passes close to edge endpoints. Vertex diffraction is necessary to compensate for such discontinuities, as shown below.

\textbf{2) Vertex diffraction}\\
Vertex diffraction accounts for scattering at points where multiple edges intersect. 
In curved surfaces discretized into facets, such vertices appear frequently.
Vertex diffraction compensates for edge diffraction discontinuities at edge endpoints, since discretization facets have finite sides/edges.  
The formulation of the vertex diffraction coefficient is given in \cite{albani2009utd} as 

\begin{equation}
    D_V \sim \sum_{i=1}^{4}  B\!\left(\Phi_i\right) T_{\mathrm{GFI}}\!\left(b, a_i\right)
\end{equation}

The cotangent term from edge diffraction is replaced by the more complex trigonometric function $B\!\left(\Phi\right)$, while the transition is governed by two parameters: $b$ is the optical distance between the vertex and edge rays, and $a$ is identical to that in the edge diffraction formulation. 
The equation involves a special function named the Generalized Fresnel Integral (GFI) $T_{\mathrm{GFI}}$. 
Its role is to ensure uniform transition to edge-diffracted and specular fields.
The structure of the equation resembles classical UTD edge diffraction and is applied to each edge constituting the vertex.

Two transition regimes can be distinguished:
\begin{itemize}
    \item Single transition: when the edge-diffracted ray approaches an edge endpoint, restoring continuity of the edge-diffracted field
    \item Double transition: when transition occurs with respect to two edges constituting the vertex, restoring continuity of the specular field
\end{itemize}

\textbf{3) Double-edge diffraction}

Double-edge diffraction models sequential diffraction at two edges. In our study, it is essential to model propagation into shadow regions that cannot be reached by single-edge diffraction. 
However, it is well known that simple cascading of ordinary UTD diffraction coefficients in \eqref{eq:ordinaryUTDcoeff} fails when the second edge lies in the transition region of the field diffracted by the first edge: to overcome this problem, a different formulation of the double-edge diffraction coefficient involving higher-order Fresnel's integrals is necessary, similarly to what has been done for vertex diffraction.
The above-mentioned formulation is presented in \cite{albani2005uniform} by two terms

\begin{equation}
    \tilde{D}_{EE} \sim \sum_{i,j=1}^{4}  \cot \!\left(\frac{\Phi_{1,i}}{2n_1}\right) \cot \!\left(\frac{\Phi_{2,j}}{2n_2}\right) \tilde{T}(a_{i}, b_{j}, w)
\end{equation}

\begin{equation}
    \tilde{\tilde{D}}_{EE} \sim \sum_{i,j=1}^{4} \csc^2\!\left(\frac{\Phi_{1,i}}{2n_1}\right) \csc^2\!\left(\frac{\Phi_{2,j}}{2n_2}\right) \tilde{\tilde{T}}(a_{i}, b_{j}, w)
\end{equation}
where the parameters are defined separately for edge 1 ($\Phi_{1}$, $n_1$, $a$) and edge 2 ($\Phi_{2}$, $n_2$, $b$); $\tilde{T}$ and $\tilde{\tilde{T}}$ are special Fresnel's functions which can be derived from the mentioned above GFIs, and $w$ is the distance parameter.

When one of the edges is close to a shadow boundary, a transition occurs, and double-edge diffraction ensures continuity of the single-edge diffracted field.
When both edges are near shadow boundaries, a double transition restores continuity of the specular field.

Terms $\tilde{D}_{EE}$ and $\tilde{\tilde{D}}_{EE}$ exhibit different asymptotic behaviour and different physical contributions.
In the configuration considered in this work, when the field propagates along a facet, $\tilde{D}_{EE}$ describes the hard diffraction component, while $\tilde{\tilde{D}}_{EE}$ represents the soft one.
The soft component is asymptotically weaker outside transition regions, but becomes of the same order near the double transition regions, and is therefore essential for achieving smooth behaviour of the specular field.

\textbf{4) Cascaded edge and vertex (EV, VE) diffraction}
The double-edge diffraction described above assumes edges of infinite length; however, in discretized meshes, mixed edge–vertex (EV) and vertex–edge (VE) interactions become significant for the shadow region modeling (and vertex-vertex interactions to a smaller extent, which are not considered in this work).

\rev{These mixed diffraction interactions are usually neglected in standard RT tools, since no exact solution is available in the literature. However, in the present work we present a heuristic solution, as described in the following. First of all, a simple cascading of the corresponding diffraction coefficients is applied as:}
\begin{equation}
    D_{EV} \sim \sum_{i,j=1}^{4}  \cot \!\left(\frac{\Phi_{1,i}}{2n_1}\right)  B\!\left(\Phi_{2,j}\right)T_{EV}(a_{i}, b, c_{j}, w)
\end{equation}
where $T_{EV}(a, b, c, w)$ is the resulting transition function.

The transition function $T_{EV}$ cannot be obtained by a simple cascading of the respective edge and vertex transition functions; they have to be appropriately adjusted.
For the EV case, let $a$ denote the edge parameter, $b$ and $c$ the vertex parameters, and $w$ the same parameter as in the double-edge formulation.
To account for transition-region behavior, the following approximations are applied:

A) Outside the transition region: $T_{EV}(a, b, c, w) = F(a) \cdot T_{\mathrm{GFI}}(c, b)$.

B) Close to the transition of the edge: $T_{EV}(a, b, c, w) = F (a) \cdot T_{\mathrm{GFI}}(c, b / \sqrt{(1-w^2)})$, which also holds for the double transition region.

Between these two regimes, a linear interpolation is applied. 
\rev{A limitation of the proposed direct chaining method is that it can only describe the hard diffraction component, so at the moment the soft component cannot be represented.}

In addition to the aforementioned limitation in the EV and VE diffraction combinations, the current double-diffraction formulation may also have numerical instabilities in configurations where the two edges involved in double-edge diffraction share the same facet and terminate at the same vertex. When the double-diffracted ray approaches the vertex, the transition to vertex diffraction is not smooth, since vertex diffraction was formulated to compensate for the discontinuity of a single-diffracted ray, not a double-diffracted one.
\rev{A rigorous treatment of this transition would require an additional uniformization mechanism \cite{maci1998itd}, which is not readily available in a form suitable for efficient implementation in practical RT frameworks.
Therefore, this interaction is treated as a limitation of the present model and remains a topic for future work.}

\setlength{\parskip}{0ex plus 0ex minus 0ex}

{\small

\phantomsection  
\addcontentsline{toc}{chapter}{Bibliography}

\bibliographystyle{IEEEtran}	

\nocite{*}		

\bibliography{references}		
}

\end{document}